\documentclass[9pt, conference]{IEEEtran}
\IEEEoverridecommandlockouts

\usepackage{cite}
\usepackage{amsmath,amssymb,amsfonts}
\usepackage{algorithmic}
\usepackage{graphicx}
\usepackage{textcomp}
\usepackage{xcolor}
\usepackage{booktabs}
\usepackage{multirow}
\usepackage{hyperref}
\usepackage{adjustbox}

\usepackage{caption}

\def\BibTeX{{\rm B\kern-.05em{\sc i\kern-.025em b}\kern-.08em
    T\kern-.1667em\lower.7ex\hbox{E}\kern-.125emX}}
\begin{document}

\title{Efficient Long-Form Speech Recognition for General Speech In-Context Learning\\
}

\author{\IEEEauthorblockN{$\text{Hao Yen}^{1,2}\textsuperscript{*}$\thanks{\textsuperscript{*}Work done as an intern at Microsoft}$, \text{Shaoshi Ling}^{1}, \text{Guoli Ye}^{1}$}
\IEEEauthorblockN{$^1$Microsoft Corporation, Redmond, WA, USA\\$^2$School of Electrical and Computer Engineering, Georgia Institute of Technology, GA, USA}
}

\maketitle

\begin{abstract}
We propose a novel approach to end-to-end automatic speech recognition (ASR) to achieve efficient speech in-context learning (SICL) for (i) long-form speech decoding, (ii) test-time speaker adaptation, and (iii) test-time contextual biasing. Specifically, we introduce an attention-based encoder-decoder (AED) model with SICL capability (referred to as SICL-AED), where the decoder utilizes an \textit{utterance-level} cross-attention to integrate information from the encoder's output efficiently, and a \textit{document-level} self-attention to learn contextual information. Evaluated on the benchmark TEDLIUM3 dataset, SICL-AED achieves an 8.64\% relative word error rate (WER) reduction compared to a baseline utterance-level AED model by leveraging previously decoded outputs as in-context examples. It also demonstrates comparable performance to conventional long-form AED systems with significantly reduced runtime and memory complexity. Additionally, we introduce an in-context fine-tuning (ICFT) technique that further enhances SICL effectiveness during inference. Experiments on speaker adaptation and contextual biasing highlight the general speech in-context learning capabilities of our system, achieving effective results with provided contexts. Without specific fine-tuning, SICL-AED matches the performance of supervised AED baselines for speaker adaptation and improves entity recall by 64\% for contextual biasing task.
\end{abstract}


\section{Introduction}
\label{sec:intro}
The rising popularity of AI-assisted devices has sparked heightened interest in Automatic Speech Recognition (ASR) for personalized user experiences~\cite{Sim2019,Huang2020}. Traditionally, ASR systems have primarily relied on supervised training methods to perform correct biasing and adaptation~\cite{Weninger2019,huang2021}. However, a common requirement to adapt a high-performance ASR system is to prepare extensive labeled training data, which is often impractical in real-world scenarios. In fact, it is generally favorable to build an unified ASR system that can perform test-time adaptation to a wide range of acoustic characteristics without requiring extensive retraining, offering a more robust and scalable solution for personalized speech recognition.

In-context learning (ICL)~\cite{Dong2023} emerges as a promising technique, particularly in the field of Large Language Models (LLMs)~\cite{Brown2020,Touvron2023}. While ICL has been widely studied in LLMs for image-text and audio-text tasks~\cite{Zhao2021,Tsimpoukelli2021,Hsu2024}, its potential in advanced automatic speech recognition (ASR)~\cite{Li2021} systems remains largely unexplored. Given the strong linguistic connection between speech and text, large-scale ASR models might be capable of leveraging contextual cues for ICL, enabling models to perform test-time adaptation based on provided examples and making this a crucial area for further research. Inspired by the success of ICL, speech in-context learning (SICL)~\cite{Pan2023,Wang2024}, which can be viewed as ICL grounded on spoken language and speech processing, has gained traction as a promising solution to address the aforementioned scenarios.

In the realm of SICL, a common requirement is to provide a sufficient amount of paired speech-text data as examples for the models to learn task-specific patterns. Consequently, ASR systems must be capable of handling long-form speech-text pairs to effectively perform SICL. An attention-based architecture~\cite{Vasawani2017} naturally offers this advantage, as it excels at capturing long-range dependencies and can process sequences in parallel. Opting for a attention-based encoder-decoder (AED) ASR~\cite{Dong2018,Zhao2019} system provides the benefit to efficiently model and utilize context over extended sequences, making it suitable for long-form ASR tasks. Despite the advantages of AED systems, the quadratic memory complexity of attention has typically restricted AED models to utterance-based speech processing, hindering their ability to effectively utilize long-form contexts.

In this work, our goal is to build an end-to-end ASR model that can leverage long-form contextual information to enable speech in-context learning (SICL) capability. The design concept of the proposed system consists of a modified attention-based encoder-decoder (AED) model, referred to as SICL-AED, where each of the decoder layer performs utterance-level cross-attention and document-level self-attention. By doing so, SICL-AED is able to efficiently integrate speech and text information between encoder and decoder by utterances, restricting each hidden vectors to look at only the relevant outputs of the encoder that correspond to the current utterances. The document-level self-attention can effectively utilize all the available contextual information by attending to all current and previous hidden vectors. To further enhance the SICL capabilities, an in-context fine-tuning (ICFT) technique is introduced, aiming to enable the model to effectively learn from the provided contextual information during inference. Our experiments demonstrate that our proposed SICL-AED not only delivers efficient long-form speech recognition by utilizing previous decoded outputs, but also achieves competitive results compared to fully supervised models in speaker adaptation task without any fine-tuning. The integration of ICFT further enhances the SICL-AED model, making it more effective in leveraging long-form contextual information for improved performance and outperforms the baseline for the contextual biasing task.

\section{Related Work}
\label{sec:related}
\vspace{-.1cm}
\subsection{Attention-based Encoder-Decoder for Long-form ASR}
\vspace{-.1cm}
Previous research has explored long-form speech processing within attention-based encoder-decoder (AED) models, employing various strategies to handle extended speech inputs. The most straightforward approach involves concatenating successive speech utterances or transcriptions. In~\cite{Hori2020,Hori2021}, the authors presented a context-expanded Transformer, extending earlier work to accelerate the decoding process within a streaming AED architecture. Chen et al.~\cite{Chen2024} replaced the classical attention mechanism of Transformers and showed that utilize entire spoken documents is possible during both training and testing. Although word error rate reductions were observed, these methods focus solely on long-form speech recognition with consecutive utterances. However, for general speech in-context learning, in-context examples are often sampled randomly, where utterances are not guaranteed to be sequential.

Another category of approaches utilize auxiliary encoders to model long-context information. Masumura et al.~\cite{Masumura2021} introduced a hierarchical text encoder combined with the distillation of large-context knowledge beyond the current utterance, enabling the model to effectively capture long-form information while preserving ASR performance. Similarly, Wei et al.~\cite{Wei2022,Wei2022_2} employed a residual attention module, and pretrained encoders to accelerate the convergence speed and well model the long-range global dependencies within each input sequence for conversational ASR. Despite the effectiveness in long-form speech recognition, these methods require additional encoder, pretrained models, and modifications to the underlying AED architecture, which adds complexity and increases the computational demands of the model. 

\vspace{-.1cm}

\subsection{Speech In-context Learning (SICL) for ASR}
\label{sec:icl}
\vspace{-.1cm}
In-context learning (ICL) refers to a model's ability to make inferences based on the patterns of in-context examples provided within the context. Recently, several studies have been conducted to incorporate LLMs with ASR models to boost the performances of standalone ASR systems. In~\cite{Chiu2021,Zheng2021,Xu2022}, the authors leveraged LLMs to rescore ASR output hypotheses, which often requires fine-tuning existing LLMs. COSMIC~\cite{Pan2023} presented a cost-effective method to integrate speech into LLM with instruction-following and in-context learning capabilities. ~\cite{Fathullah2024} extended the capabilities of LLMs by directly attaching a small audio encoder allowing it to perform speech recognition. An obvious limitation of these previous approaches is the requirement for external pre-trained LLMs, carefully curated prompts, and additional fine-tuning. In this study, we aim to equip ASR models with inherent in-context learning capabilities directly without any additional LLMs and prompt-tuning.

Unlike LLMs, which mainly deal with text modality, speech in-context-learning (SICL) requires feeding the paired speech inputs and corresponding transcriptions as the in-context examples to the encoder and decoder respectively. In~\cite{Wang2024,Bay2024}, the authors investigates the speech in-context learning abilities of the Whisper models~\cite{Radford2023} for test-time adaptation on Chinese dialects. In contrast to~\cite{Wang2024,Bay2024}, which focus on language- and speaker-level adaptation, we aim to provide a system that can support general speech in-context learning. In addition, while the Whisper model is limited to handling only 30 seconds of speech input, our proposed approach can potentially process document-level long audio, providing a greater opportunity to incorporate broader in-context examples.

\begin{figure}[t!]
\vspace{-.5cm}
    \centering
    \includegraphics[width=0.6\linewidth]{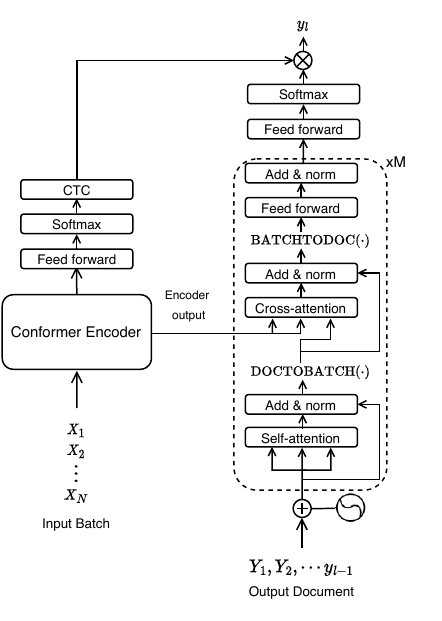}
    \vspace{-.4cm}
    \caption{The SICL-AED framework.}
    \label{fig:proposed}
    \vspace{-.4cm}
\end{figure}

\section{Proposed Method}
\vspace{-.1cm}
In the following sections, "utterance-level" refers to processing individual speech sentences or phrases. In contrast, "document-level" processing involves handling multiple utterances grouped together, providing a broader context across several sentences. 
\vspace{-.1cm}

\subsection{SICL-AED}
\label{sec:proposed}
\vspace{-.1cm}
Figure~\ref{fig:proposed} illustrates our proposed approach, SICL-AED, which builds upon a standard attention-based encoder-decoder ASR model, featuring a Conformer~\cite{Gulati2020} encoder and a modified Transformer decoder. Unlike conventional AED models, SICL-AED introduces two additional functions within each decoder layer: \textsc{DocToBatch} and \textsc{BatchToDoc}. After the self-attention layer, \textsc{DocToBatch} converts the hidden vectors from the document-level to the utterance-level batch, allowing the model to efficiently compute cross-attention with the encoder outputs on an utterance-by-utterance basis. Following cross-attention, \textsc{BatchToDoc} reverses this operation, transforming the utterance-level hidden vectors back into document-level hidden vectors. We ensure that each hidden vector focuses only on the encoder outputs corresponding to its own utterance during cross-attention. This approach provides two significant benefits. First, it prevents the decoder from being confused by unrelated utterances, which is particularly advantageous during test-time adaptation when input speech-text pairs may not be sequential or related. Second, by limiting the scope of cross-attention, our model requires less computation and memory, making it more scalable and suitable for processing longer input sequences.

Given $N$ random speech features $(X_1, X_2, \dots, \allowbreak X_N)$, we denote $X_{1:N}$ as input batch of utterances. The Conformer encoder first processes $X_{1:N}$ and generates outputs for each individual utterance. To facilitate effective long-form decoding, during training, we concatenate the transcriptions $Y_1,Y_2,\cdots,Y_N$ corresponding to each utterance to form an \textit{output document}. For simplicity, we denote the \textit{output document} as $Y_{1:N}$. The training objective of the attention-based decoder is approximately computed from the sequence posterior $p_{\text{att}}(Y_{1:N}\mid X_{1:N})$.

\begin{equation}
\label{eq:1}
\begin{aligned}
p_{\text{att}}(Y_{1:N}\mid X_{1:N}) \approx \prod_{l=1}^L p\left(y_{l} \mid y_{1: l-1}, X_{1: N}\right),
\end{aligned}
\vspace{-.1cm}
\end{equation} where $y_{1:L}$ denotes the ground truth token sequence $(y_{1}, \dots, y_{L})$ of $Y_{1:N}$. We also incorporate an auxiliary CTC~\cite{Graves2006,Graves2014} sequence probability $p_{\text{ctc}}(\bar{Y}_{1:N}\mid X_{1:N})$, where $\bar{Y}_{1:N}$ is the ground truth token sequence processed in batch This is used in the hybrid CTC/Attention approach~\cite{Watanabe2017,Kim2017} to optimize the multi-objective loss $\mathcal{L}$ as follows:


\begin{equation}
\label{eq:2}
\begin{aligned}
\mathcal{L}=\lambda & \log p_{\text{ctc}}\left(\bar{Y}_{1:N} \mid X_{1:N}\right) \\
& +(1-\lambda) \log p_{\text{att}}\left(Y_{1:N} \mid X_{1: N}\right)
\end{aligned}
\end{equation} where $\lambda$ is a scaling factor to balance the CTC and attention scores. Since our proposed SICL-AED can process multiple utterances at once, it inherently supports SICL. This capability allows the model to predict output tokens for a target utterance by leveraging previously processed speech-text pairs as contextual examples, enhancing its adaptability to varying contexts.

\vspace{-.1cm}

\subsection{Speech In-Context Learning (SICL) Capability}
\label{sec:sicl}
\vspace{-.1cm}

\subsubsection{General Long-form Decoding}
\label{sec:long}
Long-form decoding is a specific application of SICL where the model decodes an entire sequence of utterances while continually using previously decoded utterances as in-context examples. For a given utterance $X_N$, the decoder aims to find the most probable token sequence $\hat{Y}_N$, conditioned on the input batch $X_{1:N}$ and the previously decoded sequence $\hat{Y}_{1:N-1}$. Here, $X_{1:N-1}$ represents the input context of earlier utterances and $\hat{Y}_{1:N-1}$ provides the corresponding output context generated by the model. The decoder utilizes these in-context examples to refine its prediction for $\hat{Y}_N$, formulated as follows:

\begin{equation}
\label{eq:3}
\vspace{-.1cm}
\begin{aligned}
\hat{Y}_N= & \underset{Y_N \in \mathcal{U}^*}{\operatorname{argmax}}\{\lambda \log p_{\text{ctc}}(Y_N \mid X_{N}) \\
& +(1-\lambda) \log p_{\text{att}}(Y_N \mid \hat{Y}_{1: N-1}, X_{1: N})\}
\end{aligned}
\end{equation} where $\mathcal{U}$ is the vocabulary.

The decoding process starts with the first utterance, $X_1$, processed independently without any prior context, generating the initial hypothesis $\hat{Y}_1$. For the next utterance, $X_2$, the system utilizes both the input batch $X_{1:2}$ and the previously decoded output $\hat{Y}_1$ to decode $X_2$ more accurately, producing the hypothesis $\hat{Y}_2$. This sequential decoding continues for subsequent utterances, allowing the model to leverage context effectively across the entire long-form speech input.


\subsubsection{Test-time Speaker Adaptation}
\label{sec:sa}
In contrast to long-form decoding, where in-context examples are derived from previously generated outputs, test-time speaker adaptation leverages in-context examples obtained directly from ground truth data. This process begins by randomly sampling $N-1$ speech-text pairs, $X_{1:N-1}$ and $Y_{1:N-1}$, from a specific speaker. These sampled pairs are combined with the target utterance $X_N$ to form the input batch $X_{1:N}$, which is then fed into the model. The objective remains similar to that in Eq.~\ref{eq:3}; however, instead of conditioning on previous decoded hypotheses $\hat{Y}_{1:N-1}$, the decoder relies on the ground truth transcriptions $Y_{1:N-1}$. This allows the model to adapt more effectively to a specific speaker by directly utilizing accurate speech-text pairs as contexts, thereby improving recognition performance for the target utterance.

\subsubsection{Contextual Biasing}
\label{sec:rec}
Traditional approaches to ASR biasing \cite{sathyendra2022contextual, fu2023robust, tang2024improving, Yen2022} are limited to text-form phrases, making it challenging to accurately bias for non-standard pronunciations, such as foreign names. Our method addresses this issue by leveraging acoustic information from in-context audio examples. This allows our model to learn correct pronunciation and spelling mappings, even for words from other languages. This process is similar to speaker adaptation in section \ref{sec:sa}, where the model uses audio-text pair examples that include the biasing phrases and combines them with the target utterance. As a result, the model can produce accurate transcription outputs by referencing the in-context examples.

\subsection{In-context Fine-tuning (ICFT)}
\label{sec:icft}
\vspace{-.1cm}
As described in Section~\ref{sec:proposed} and Eq.~\ref{eq:2}, our SICL-AED model is trained with standard ASR loss. Without an explicit objective to encourage the model to utilize previous contexts, there is no guarantee that it will effectively perform SICL during inference. To address this, we introduce an In-context Fine-Tuning (ICFT) technique as an auxiliary objective during fine-tuning to ensure the model focuses on previous contexts or in-context examples. During this fine-tuning stage, we keep the model structure unchanged. For each target sequence, we incorporate a few speech-text pairs from the same speaker as the in-context examples. We then randomly select a word that appears in both the target sequence and these examples, and randomly modify this word by inserting, deleting, or substituting a few letters. The model is then tasked to predict the modified word instead of its original form. The objective is the same as \ref{eq:2} but it is applied only to the target sequence containing the modified word.

\begin{table}[t!]
\centering
\caption{Testing WER (\%) of the long-form decoding results. Utt. and Doc. indicate whether the training was conducted on utterances or document-level audio, respectively. Models marked with $(^\ast)$ are results taken from the corresponding papers.}
\vspace{-.1cm}
\label{tab:asr_results}
\begin{tabular}{llcc}
\toprule
\multirow{2}{*}{\textbf{Model}} & \multirow{2}{*}{\textbf{Train (Dur.)}} & \multicolumn{2}{c}{\textbf{WER}($\downarrow$)} \\
&   & \multicolumn{1}{c}{Utt.} & \multicolumn{1}{c}{Doc.} \\ \midrule \midrule
Transformer$^\ast$~\cite{Hori2020} & Doc. & -- & 8.1 \\
Fast-Conformer$^\ast$~\cite{Rekesh2023} & Doc. ($>30$ mins) & --  & 7.5   \\ 
Flash Attention$^\ast$~\cite{Chen2024} & Doc. ($>30$ mins) & --  & \textbf{7.2}   \\  \midrule
\multirow{2}{*}{AED} & Utt. & 9.03  & 100.23 \\
                     & Doc. (3 mins) & 9.29  & \textbf{8.15}   \\  \midrule \midrule
SICL-AED & Doc. (3 mins) & \textbf{8.85} & 8.25\\ \bottomrule
\vspace{-.4cm}
\end{tabular}
\end{table}

\begin{table}[t!]
\centering
\caption{Inference Forward Pass Time and GPU Memory Usage}
\vspace{-.1cm}
\label{tab:mem}
\begin{tabular}{llcc}
\toprule
\textbf{Model} & \textbf{30s} & \textbf{90s} & \textbf{180s} \\ \midrule
AED & 5s/1.5G & 21s/5G & 74s/12G \\ \midrule
SICL-AED& 5s/1.5G & \textbf{14s/2.5G} & \textbf{31s/4.5G} \\ 
\bottomrule
\vspace{-.4cm}
\end{tabular}
\end{table}

\section{Experiments and Results}
\label{sec:exp}
\vspace{-.1cm}

\subsection{Datasets \& Experimental Settings}
We evaluate ASR models on the TEDLIUM3~\cite{Hernandez2018} dataset, which is composed of crawled TED talks. Since TEDLIUM3 is widely used for evaluating long-form speech recognition~\cite{Chen2024,Hori2020}, we chose to retain this dataset for the speaker adaptation experiments to ensure consistency and maintain an in-domain setting. However, because TEDLIUM3 is not specifically designed for speaker adaptation tasks, we randomly selected 5 minutes of speech per speaker from the test set to serve as training set, which can also be used as in-context examples. The remainder is considered testing data for speaker adaptation task, resulting in a total of 768 utterances. For contextual biasing experiments, we use an internal test set which contains 20 names across 120 utterances.

The Conformer encoder contains two convolutional layers that subsample the time frame by a factor of 4, followed by 18 conformer layers. Each conformer layer has a multi-head attention with 8 heads, and a depth-wise convolution with kernel size of 3. The multi-head attention and the depth-wise convolution are sandwiched between GLU-variant-based~\cite{Shazeer2020} feedforward layers with dimension of 684. The decoder consists of 6 layers , with 2048-dim feedforward layer as well. The embedding dimension is set to be 512 for both encoder and decoder. All the ASR models are trained using a hybrid CTC/attention~\cite{Watanabe2017} loss approach, with a CTC weight set to 0.2. We conduct experiments on up to 3-minutes speech. However, our system can naturally support longer inputs and scalable with the increase of memory size. For all document-level training, models are pre-trained on utterance-level ASR before being fine-tuned on the document-level audio. In-context Fine-tuning (ICFT) is performed after SICL-AED is well trained.

\vspace{-.1cm}

\subsection{Long-form ASR Results}
Table~\ref{tab:asr_results} presents the results for long-form ASR on the TEDLIUM3 dataset, comparing the performance of our proposed SICL-AED model with the conventional AED model and prior works, such as Flash Attention~\cite{Chen2024} and Fast-Conformer~\cite{Rekesh2023}, for both utterance-level and document-level training. The document-level AED model, Flash Attention~\cite{Dao2022,Chen2024}, and Fast-Conformer~\cite{Rekesh2023} models are trained by concatenating utterances of varying lengths to form document-level inputs. While Flash Attention and Fast-Conformer are trained on document-level audio exceeding 30 minutes, our AED models work with inputs up to 3 minutes, fitting within our memory capacity. When evaluated on utterances, the SICL-AED model, which is trained exclusively on 3-minute inputs, achieves a WER of 8.85\%, slightly outperforming the utterance-level AED model (9.03\% to 8.85\%). The slight reduction in WER may be due to the random sampling of input batch, allowing the model learn more patterns and thus generalize better to the test set. This result is particularly compelling given that the SICL-AED model does not suffer from the overfitting observed in the document-level AED model, where the WER increases from 9.03\% to 9.29\%. This indicates that the utterance-level cross-attention mechanism in SICL-AED effectively focuses on relevant contextual information without over-relying on extended context windows, thereby enhancing its generalization ability across different input lengths.

Moreover, when evaluated on 3-minute document-level audio, the SICL-AED model achieves a WER of 8.25\%, outperforming the utterance-level AED model with 8.62\% relative word error rate reduction (WERR), and closely matching the document-level AED model's performance of 8.15\%. This demonstrates that the SICL-AED model effectively leverages the long-form context. Compared to prior works, such as Flash Attention and Fast-Conformer, there are performance differences when evaluated on document-level audio. These gaps may be attributed to differences in model architectures, tokenizations, and the use of much longer document-level audio in Flash Attention and Fast-Conformer, where both models can process over 30 minutes of audio. Additionally, the slight difference in SICL-AED's performance compared to the document-level AED model is not unexpected, given that SICL-AED is specifically designed to balance efficiency with flexibility in handling both consecutive and non-consecutive utterances. Unlike the document-level AED model, which can scan entire encoder outputs composed of consecutive utterances and benefit from rich contextual information, SICL-AED is optimized for general speech in-context learning. This includes scenarios where utterances are not guaranteed to be sequential and often lack such dense context and relationship. Therefore, while the document-level AED model may gain a marginal advantage in strictly sequential inputs, SICL-AED's design prioritizes robustness and adaptability across varying and potentially disjoint utterances, making it a more versatile solution. Moreover, as shown in Table~\ref{tab:mem}, SICL-AED can achieve significantly faster runtimes (33.3\% and 58.1\% reduction in forward pass time), and reduced memory complexity (50\% and 62.5\% reduction in memory usage) for 90-second and 180-second audio. In addition, it is worth mentioning that our proposed SICL-AED can be trained using only 32GB Nvidia V100 GPUs while other conventional AED models require larger memory capacities. This highlights its practicality for real-world applications where computational efficiency is crucial.

\begin{table}[t!]
\centering
\caption{Testing WER (\%) of the speaker adaptation results. $\mathcal{C}$ refers to the types of in-context examples, including text and speech.}
\vspace{-.1cm}
\label{tab:sa_results}
\begin{adjustbox}{width=0.66\columnwidth}
\begin{tabular}{llcc}
\toprule
\textbf{Model} & \textbf{Train set} & $\mathcal{C}$ & \textbf{WER($\downarrow$)} \\ \midrule \midrule
AED & Utt. & -- & 8.43 \\ \midrule
AED-FT & Utt. & -- & 8.11 \\\midrule \midrule
\multirow{2}{*}{SICL-AED} & \multirow{2}{*}{Doc.} & -- & 8.40 \\ 
& & text  & 8.57 \\ \midrule
SICL-AED & \multirow{2}{*}{Doc.} & \multirow{2}{*}{speech+text}  & 8.20 \\
+ ICFT & & & \textbf{8.13} \\ \bottomrule
\vspace{-.6cm}
\end{tabular}
\end{adjustbox}
\end{table}

\vspace{-.1cm}

\subsection{Speaker Adaptation Results}
\vspace{-.1cm}
Table~\ref{tab:sa_results} presents the results for speaker adaptation. The TED-FT model represents 11 speaker-dependent systems that were fine-tuned separately for each speaker and the results are average of the 11 models tested on individual speakers. Two types of in-context examples are investigated in this work, including speech and text. A text-only SICL-AED refers to models trained and tested to predict the target utterance by providing only transcriptions as previous in-context examples.  

Compared to the AED model without fine-tuning, AED-FT shows a modest improvement in WER, reducing it from 8.43\% to 8.11\%. When examining the performance of text-only SICL-AED, we observe a slight increase in WER compared to the utterance-level AED model (8.57\% vs. 8.43\%). This suggests that relying solely on randomly selected text-based contexts, such as transcriptions, does not provide improvement in the model's ability to capture specific speaker characteristics without the inclusion of audio information. However, when both speech and text are provided as in-context examples, our system outperforms the utterance-level AED model, with 8.20\% WER, indicating that speech-text pairs are crucial to learn useful contextual information for SICL. Moreover, when combined ICFT with SICL-AED, our system achieves a WER of 8.13\%, demonstrating comparable performance compared to TED-FT. This finding suggests that in limited-resource scenarios, SICL-AED may offer a more effective and efficient approach for test-time speaker adaptation. That is, we do not have to fine-tune and store 11 speaker-specific models and the adaptation is being done \textit{on-the-fly}, making it a promising solution for real-world applications for large-scale products with millions of speakers and parameters.

\begin{table}[t!]
\centering
\caption{Entity Recall (\%) of the Contextual Biasing Results.}
\vspace{-.1cm}
\label{tab:rt}
\begin{adjustbox}{width=0.66\columnwidth}
\begin{tabular}{llcc}
\toprule
\textbf{Model} & \textbf{Train set} & \textbf{WER($\downarrow$)} & \textbf{Recall($\uparrow$)} \\ \midrule \midrule
AED & Utt. & 34.08 & 12.82 \\ \midrule
SICL-AED & Doc.&  33.99 & 12.85 \\ 
+ICFT & Doc.& \textbf{33.73} & \textbf{21.03} \\
\bottomrule
\vspace{-.6cm}
\end{tabular}
\end{adjustbox}
\end{table}

\vspace{-.1cm}

\subsection{Contextual Biasing Results}
\vspace{-.1cm}
Table~\ref{tab:rt} presents the entity recall results for contextual biasing. The utterance-level AED model, which lacks specific knowledge about rare or unseen entities, shows low recall performance at 12.82\%, as most of these names never appeared in the training data. For the SICL-AED model, we provided 20 utterances containing 20 names as in-context examples during inference. However, SICL-AED only slightly improved to 12.85\% recall, suggesting that it could not effectively leverage the limited contextual information from these examples. In contrast, when combining SICL-AED with the in-context fine-tuning (ICFT) technique, we observe a substantial improvement, achieving an entity recall of 21.03\% without any regression in general WER. This result represents a 64\% relative increase compared to the utterance-level AED baseline, demonstrating the strong capability of ICFT to enhance SICL performance by allowing the model to adapt more effectively to previous contexts. 

\vspace{-.1cm}

\section{Conclusion}
\label{sec:conclusion}
In this work, we introduce \textbf{SICL-AED}, an efficient attention-based encoder-decoder ASR model designed for long-form speech recognition with speech in-context learning (SICL) capabilities. By employing \textit{utterance-level} cross-attention and \textit{document-level} self-attention, SICL-AED leverages contextual information to enhance both recognition accuracy and computational efficiency. Our model achieves an 8.64\% relative WER reduction compared to utterance-level AED models and demonstrates comparable performance to traditional document-level AED model, with significant reductions in runtime and memory usage. Additionally, incorporating in-context fine-tuning (ICFT) further improves performance in speaker adaptation and contextual biasing tasks, highlighting SICL-AED’s potential for general SICL and real-world ASR applications. In future work, we will focus on extending these capabilities to more challenging adaptation scenarios with speech in-context learning.

\clearpage
\bibliographystyle{IEEEtran}
\bibliography{refs}

\end{document}